# A Fitting Model for Asymmetric I-V Characteristics of Graphene Field-Effect Transistors for Extraction of Intrinsic Mobilities

Akira Satou, *Senior Member, IEEE*, Gen Tamamushi, Kenta Sugawara, Junki Mitsushio, Victor Ryzhii, *Fellow, IEEE*, and Taiichi Otsuji, *Fellow, IEEE*

*Abstract*—A fitting model is developed for accounting the asymmetric ambipolarities in the I-V characteristics of graphene field-effect transistors (G-FETs) with doped channels, originating from the thermionic emission and interband tunneling at the junctions between the gated and access regions. Using the model, the gate-voltage-dependent intrinsic mobility as well as other intrinsic and extrinsic device parameters can be extracted. We apply it to a top-gated G-FET with a graphene channel grown on a SiC substrate and with SiN gate dielectric that we reported previously, and we demonstrate that it can excellently fit its asymmetric I-V characteristic.

## I. Introduction

GRAPHENE field-effect transistors (G-FETs) have been vastly investigated for ultrahigh speed electronics [1] since the discovery of graphene [2,3]. The demonstration of intrinsic carrier mobility higher than 200,000 cm$^2$/Vs of peeling graphene at room temperature [4], which was extracted by excluding all the scattering mechanisms other than acoustic-phonon scattering, stimulated the development of fabrication technologies of G-FETs that maintain the ultimately high mobilities and realize large-scale integration of them at the same time. So far, an intrinsic cutoff frequency above 300 GHz has been demonstrated [5] by G-FETs based on either epitaxial graphene grown on silicon carbide (SiC) or on graphene grown by chemical vapor deposition (CVD).

Up to now, techniques such as utilization of C-face of 4H- or 6H-SiC [6] and epitaxial CVD growth on hetero-epitaxial metal substrates [7] have been developed to realize high-quality, large-area graphene, while techniques such as hydrogen intercalation between graphene and SiC substrates [8,9], encapsulation of graphene by hexagonal boron nitride layers for CVD or peeling graphene [10], and low-damage deposition of silicon nitride (SiN) as gate dielectrics [11] have been introduced to suppress interfacial effects of substrates and gate dielectrics. Deposition of metal contacts to reduce contact resistances as well as control of doping to graphene are of equal importance on the device performances.

For evaluation of growth and gate-stack technologies of G-FETs on their performances, the most important figure of merit is the intrinsic mobility, i.e., the mobility in the gated region of a G-FET. It is directly influenced by carrier scattering mechanisms originated from scatterers in graphene, substrates, and gate dielectrics. There exist several models [12-15] which can be used for fitting the I-V characteristics of G-FETs and extracting the intrinsic mobilities. However, they assume gate-voltage-independent mobilities and, more importantly, do not take into account the asymmetric ambipolarities in the I-V characteristics of G-FETs with doped channels, which arises due to the thermionic emission and interband tunneling between the gated and access regions [16, 17]. To fully evaluate and understand the device performance from the I-V characteristics, the gate-voltage-dependent intrinsic mobilities and the asymmetry must be taken into account. Especially, the latter is important to extract the hole mobility correctly; otherwise, it causes an unphysically large discrepancy in the electron and hole mobilities (see, for example, [11]).

The purpose of this paper is to develop a fitting model to correctly extract the gate-voltage-dependent intrinsic mobilities in G-FETs and other intrinsic and extrinsic device parameters, taking into account the asymmetric ambipolarities in the I-V characteristics of doped channels, and to demonstrate its applicability to real devices. The model includes the following factors: (i) energy-dependent momentum relaxation time, which results in the gate-voltage-dependent Drude conductivity and, hence, the gate-voltage-dependent mobility, and (ii) the thermionic emission and interband tunneling at *n-i/i-n* or *n-p/p-n* junctions formed at the edges of the gated region, which results in the asymmetric ambipolarities in the I-V characteristics. We verify the model by applying it to a top-gated G-FET with a graphene channel grown on a SiC substrate and with SiN gate dielectric. We demonstrate that our model can fit the measured asymmetric I-V characteristic excellently and can extract the gate-voltage-dependent intrinsic mobility and other intrinsic and extrinsic device parameters. At the same time, it is turned out that there is a fundamental

Manuscript received xxx xx, 2016; revised xxx xx, 2016; accepted xxx xx, 2016. Data of publication xxx xx, 2016; date of current version xxx xx, 2016. This work was financially supported by JSPS Grant-in-Aid for Specially Promoted Research (#23000008).

The authors are with Research Institute of Electrical Communication, Tohoku University, Sendai 980-8577, Japan (e-mail: a-satou@riec.tohoku.ac.jp).

difficulty of unique extraction of the intrinsic mobility of a G-FET from a single I-V characteristic only, because of the inseparability of the extrinsic resistance (the sum of access and contact resistances) to the gate-voltage-independent part of the resistance of the gated region. Instead, we introduce a method to find the upper and lower bounds of the mobility. Also, it is revealed that the Dirac voltage does not correspond to the dip of the I-V characteristic because of the thermionic emission and interband tunneling at the junctions.

## II. Fitting Model

In the model developed below, we restrict ourselves in the case (i) where the graphene channel with zero gate voltage is uniformly doped (either $n$ or $p$), (ii) where the G-FET operates in the linear regime, i.e., where the drain voltage is sufficiently small so that the drain current is proportional to the drain voltage, (iii) where the gated region of the channel is sufficiently long so that the Drude model for its conductivity is applicable, and (iv) where the access regions are sufficiently long so that the potential distribution around $n$-$i$/$i$-$n$ or $n$-$p$/$p$-$n$ junctions does not change by the presence of contacts. In addition, we neglect the variation of the concentration in the gated region by the voltage drop across it when applying the drain voltage, which was indicated in [18]. This together with the assumption (iii) make the calculation of the conductance in the gated region somewhat inaccurate in the proximity of the Dirac voltage, where the concentration variation is large and even charge polarity might change. Away from the Dirac voltage, however, this effect should be negligibly small as long as the source-drain voltage is small, which is guaranteed by the assumption (ii), and it should not affect the fitting of an entire I-V characteristic. Here, we consider the $n$-doped channel; the model can be easily extended to the case of the $p$-doped channel by exploiting the symmetry of the electron and hole transports. We also focus on a top-gated G-FET, as the extension to a top- and/or back-gated G-FET is straightforward, except that we need to take care of variation of the contact resistance by the back gate [19, 20].

A schematic view of a top-gated G-FET under consideration is depicted in Fig. 1(a). The following geometrical parameters are assumed to be known from the device design: the gate length $L_g$, thickness $t_g$ and relative permittivity $\varepsilon_t$ of the gate dielectric, the channel width $W_{ch}$, relative permittivity of the substrate $\varepsilon_b$, and length of the access regions $L_a$. The doping concentration, which is represented through the Fermi level, $\varepsilon_{dope}$, is to be determined by the fitting. Figures 1(b)-(e) show band diagrams in the channel at zero drain voltage with different gate voltages. Depending on the value of the gate voltage, the carrier type of the gated region becomes either $n$ (Figs. 1(b) and (c)), $i$ (Fig. 1(d)), and $p$ (Fig. 1(e)).

In this model, we take into account the electron thermionic emission and interband tunneling either at $n$-$i$/$i$-$n$ or $n$-$p$/$p$-$n$ junctions formed at the edges of the gated region (see Fig. 1 (f)). As shown in Fig. 1(g), the channel resistance can be represented as the sum of the resistance of the gated region, $R_g$, the resistance due to the thermionic emission and interband tunneling at the junctions, $R_{tt}$, and the sum of the contact resistances and access resistances, $R_s = 2(R_a+R_c)$, which do not depend on the gate voltage:

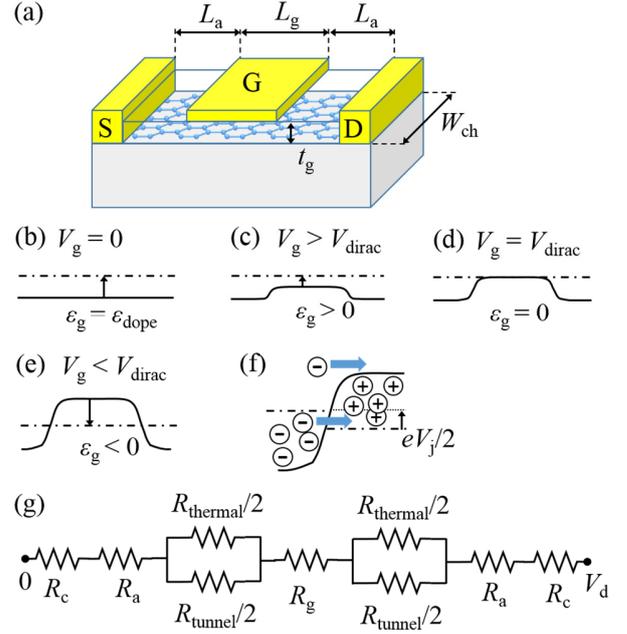

Fig. 1. (a) Schematic view of a G-FET, (b)-(e) band diagrams for different gate voltages at zero drain voltage, (f) schematic view of thermionic emission and interband tunneling at the $n$-$p$ junction between the access and gated regions with finite "junction" voltage, $V_j$, dropped at the junctions, and (g) an equivalent circuit of the G-FET (for DC characteristics).

$$R = R_g(V_g) + R_{tt}(V_g) + R_s = \frac{1}{G_g} + \frac{1}{G_{thermal} + G_{tunnel}} + R_s, \quad (1)$$

where $G_g$ is the conductance of the gated region, and $G_{thermal}$ and $G_{tunnel}$ are the conductances associated with the thermionic emission and interband tunneling at the two junctions, respectively. Expressions of $G_g$, $G_{thermal}$, and $G_{tunnel}$ shall be derived and discussed in details in the following subsections.

### A. Conductance of Gated Region

The conductance of the gated region can be expressed as

$$G_g = \frac{W_{ch}}{L_g}(\sigma_e + \sigma_h), \quad (2)$$

where $W_{ch}$ is the width of the channel, $L_g$ is the gate length, and $\sigma_e$ and $\sigma_h$ are the Drude conductivities of electrons and holes, respectively, which can be derived from the Boltzmann equation:

$$\sigma_i = \frac{e^2}{\pi\hbar^2}\int_0^\infty d\varepsilon\,\varepsilon\,\tau_i(\varepsilon)\left(-\frac{df_i}{d\varepsilon}\right). \quad (3)$$

In (3), $e$ is the elementary charge, $\hbar$ is the reduced Planck constant, $f_i = f_i(\varepsilon, \varepsilon_i, T) = \{1+\exp[(\varepsilon - s_i\varepsilon_g)/k_BT]\}^{-1}$ ($i$ = e, h, $s_e$ =

+1, and $s_h$ = -1) is the Fermi distribution function, which contains the Fermi level of the gated region, $\varepsilon_g$, and the carrier temperature, $T$, $k_B$ is the Boltzmann constant, and $\tau_i = \tau_i(\varepsilon)$ is the momentum relaxation time. Here, it is important to account for the energy-dependent relaxation time in order to accomplish a good fitting. This is because the Fermi level in graphene widely varies by the gate voltage, unlike usual semiconductors where the averaged relaxation time is reasonable.

The Fermi level of the gated region $\varepsilon_g$ is determined by the gate voltage, $V_g$, as well as the Fermi level by the doping (i.e., $\varepsilon_g$ at zero gate voltage), $\varepsilon_{dope}$. First, the Fermi level by doping $\varepsilon_{dope}$ can be obtained from the relation of the electron concentration and the gate voltage at the Dirac voltage (see Fig. 1 (d)), $V_{dirac}$,

$$c_g\left(V_{dirac} + \varepsilon_{dope}/e\right) + e\left(-n_{dope} + p_{dope}\right) = 0. \quad (4)$$

Here, $c_g = \varepsilon_t\varepsilon_0/t_g$, where $\varepsilon_0$ is the vacuum permittivity, is the static capacitance per unit area between the gate and the channel, and $n_{dope} = n(\varepsilon_{dope}, T)$ and $p_{dope} = p(\varepsilon_{dope}, T)$ are the electron and hole concentrations by doping, which are given by

$$\begin{bmatrix}n(\varepsilon_F,T)\\p(\varepsilon_F,T)\end{bmatrix} = \frac{2}{\pi\hbar^2 v_F^2}\int_0^\infty d\varepsilon\,\varepsilon\begin{bmatrix}f_e(\varepsilon,\varepsilon_F,T)\\f_h(\varepsilon,\varepsilon_F,T)\end{bmatrix}, \quad (5)$$

where $v_F$ is the Fermi velocity of graphene. Then, the Fermi level of the gated region $\varepsilon_g$ can be calculated for an arbitrary gate voltage from the relation [21]:

$$c_g\left[V_g + (\varepsilon_{dope} - \varepsilon_g)/e\right] + e\left(-n_{dope} + p_{dope}\right) \\ = e\left[n(\varepsilon_{dope} - \varepsilon_g, T) - p(\varepsilon_{dope} - \varepsilon_g, T)\right]. \quad (6)$$

Equations (4)-(6) account for the quantum capacitance of graphene, so that the Fermi level $\varepsilon_g$ around the Dirac voltage is calculated more accurately than accounting only for the static capacitance. The only unknown quantity here is the Dirac voltage, and it should be calculated as a fitting parameter.

In general, the energy-dependent momentum relaxation time $\tau_i$ should include contributions from possible scatterers: acoustic/optical phonons in graphene, disorders (short-range point defects and long-range inhomogeneities), charged impurities, grain boundaries, surface optical phonons and surface roughnesses in the substrate and the gate dielectric. To demonstrate the importance of including the energy-dependent momentum relaxation time for extraction of the intrinsic mobility, however, we focus only on two contributions in this paper and represent the momentum relaxation time as follows:

$$\frac{1}{\tau_i(\varepsilon)} = \frac{1}{\tau_{lin}(\varepsilon)} + \frac{1}{\tau_{inh}(\varepsilon)}. \quad (7)$$

The first term in (7) is the contribution from scatterers whose scattering rates are linear in energy (e.g., acoustic phonons at not too low temperatures [22] and point defects [23]):

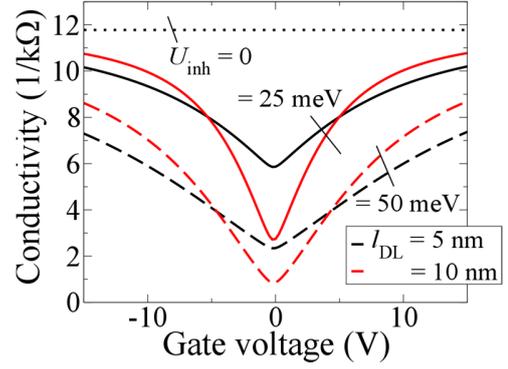

Fig. 2. Gate-voltage dependence of total conductivity, $\sigma_e+\sigma_h$, with different values of $l_{inh}$ and $U_{inh}$ for the inhomogeneity scattering, $\alpha_{lin}$ = 6.58×10⁻³ for the acoustic-phonon scattering, the Dirac voltage $V_{dirac}$ = 0, the thickness and relative permittivity of the gate dielectric $t_g$ = 45 nm and $\varepsilon_t$ = 4.2, respectively.

$$\frac{1}{\tau_{lin}} = \alpha_{lin}\frac{\varepsilon}{\hbar}, \quad (8)$$

where $\alpha_{lin}$ is a dimensionless fitting parameter; its minimum value at room temperature is bounded by the acoustic phonon scattering and is 6.58×10⁻³. The second term is the contribution from long-range inhomogeneities [23] with nonlinear scattering rate. To be specific, it includes ripples, surface roughnesses in the substrate and the gate dielectric, and, in some sense, grain boundaries if they are sufficiently dense. It is expressed as

$$\frac{1}{\tau_{inh}} = \frac{\pi}{4}\left(\frac{U_{inh}l_{inh}}{\hbar v_F}\right)^2\frac{\varepsilon}{\hbar}\Psi\left(\frac{\varepsilon l_{inh}}{\hbar v_F}\right), \quad (9)$$

where $l_{inh}$ and $U_{inh}$ are the correlation length and the characteristic potential of inhomogeneities, $\Psi(z)$ = $\exp(-z^2/2)I_1(z^2/2)/z^2$, and $I_1$ is the first-order modified Bessel function of the first kind.

The inclusion of the second term, which is nonlinear in energy, is crucial to describe the gate-voltage dependence of the conductance and, hence, of the resistance because the conductivity becomes independent of the gate voltage if the nonlinear term is absent. This behavior was already pointed out in [4, 21]. Conversely, the second term is sufficient to qualitatively describe the usual gate-voltage dependence of the conductivity, i.e., the ambipolar curve with minimum conductivity. This is clearly illustrated in Figure 2, which show gate-voltage dependences of total conductivities, $\sigma_e+\sigma_h$. In Fig. 2, we used the following parameters: $l_{inh}$ and $U_{inh}$ for the inhomogeneity scattering, $\alpha_{lin}$ = 6.58×10⁻³ for the acoustic-phonon scattering, the Dirac voltage $V_{dirac}$ = 0, the thickness of the gate dielectric $t_g$ = 45 nm, and the relative permittivity of the gate dielectric $\varepsilon_t$ = 4.2 (a measured value of SiN for our device, see Sec. III for the detail). As expected, the conductivity is constant in the case where $U_{inh}$ is equal to zero. Also, it is seen from Fig. 2 that the sharpness of the minimum changes with $l_{inh}$.

## B. Thermionic-Emission and Interband Tunneling Conductance

The conductances associated with the thermionic emission, $G_{\text{thermal}}$, and with the interband tunneling, $G_{\text{tunnel}}$, at the $n$-$i$/$i$-$n$ or $n$-$p$/$p$-$n$ junctions can be approximated analytically as follows [16,17]:

$$G_{\text{thermal}} = \alpha_{\text{thermal}} \frac{2e^2 W_{\text{ch}}}{\pi^2 \hbar^2 v_F} k_B T \log\left(1 + e^{\varepsilon_g / k_B T}\right), \quad (9)$$

$$G_{\text{tunnel}} = \begin{cases} \alpha_{\text{tunnel}} \dfrac{2e^2 W_{\text{ch}}}{\pi^2 \hbar} \sqrt{\dfrac{eE_{\text{tunnel}}}{\hbar v_F}} & (\varepsilon_g < 0), \\ 0 & (\varepsilon_g \geq 0). \end{cases} \quad (10)$$

Here, $E_{\text{tunnel}} = E_{\text{tunnel}}(V_g, V_{\text{dirac}})$ is the in-plane electric field at the "tunneling point" between the gated and access regions, where the Fermi level is equal to zero, and $\alpha_{\text{thermal}}$ and $\alpha_{\text{tunnel}}$ are dimensionless fitting parameters. If the analytical expressions work perfectly, then $\alpha_{\text{thermal}} = \alpha_{\text{tunnel}} = 1$. Equation (10) is slightly different from that derived in [17]; it reflects the fact that the doping concentrations in both source and drain sides of the access regions in top-gated G-FETs do not depend on the drain voltage. The fitting parameters $\alpha_{\text{thermal}}$ and $\alpha_{\text{tunnel}}$ are introduced to compensate to some extent for cases where the analytical expressions above do not provide good approximations. Those especially include the case of a low doping concentration and/or the case close to the Dirac voltage, where the thermal spread of the distribution functions should be taken into account for the interband-tunneling conductance, and the case where voltage drops at the junctions and the gated region change the barrier height from $\varepsilon_g$ in the thermionic-emission conductance (9).

Rather than searching for an analytical approximation of $E_{\text{tunnel}}$ for top-gated G-FETs similar to that derived in [16] for top- and back-gated G-FETs, we numerically solve the two-dimensional self-consistent Poisson equation to calculate $E_{\text{tunnel}}$. This allows us to obtain $E_{\text{tunnel}}$ in wide ranges of $V_g$ and $V_{\text{dirac}}$, as well as with different values of $\varepsilon_b$ and $\varepsilon_t$. Figure 3(a) shows $E_{\text{tunnel}}$ vs the gate voltage with different Dirac voltages (other parameters same as in Fig. 2). As the gate voltage sweeps negatively, it increases rapidly below the Dirac voltage. After that, $E_{\text{tunnel}}$ gradually decreases. This is associated with the gate fringe effect that gradually extends the $p$-region under the gate with a gradual decrease in the slope of the $n$-$p$ junction, i.e., the electric field. The decrease is less pronounced for higher Dirac voltage because the higher electron concentration in the $n$-regions screens the fringe electric field of the gate more effectively and prevents the extension of the $p$-region.

Figure 3(b) shows the gate-voltage dependence of the resistance $R_{tt}$, which is a parallel connection of $1/G_{\text{thermal}}$ and $1/G_{\text{tunnel}}$. As can be understood from Fig. 3(b) as well as (9) and (10) and Figs. 1(b)-(e), it exhibits an abrupt increase as the gate voltage sweeps negatively across the Dirac voltage. For $V_g > V_{\text{dirac}}$, $G_{\text{thermal}}$ is large and the resistance is fairly small (corresponding to Figs. 1(b) and (c)). After $V_g$ passes through $V_{\text{dirac}}$, $G_{\text{thermal}}$ exhibits exponentially decrease, so that the total resistance rises abruptly. Meanwhile, for $V_g < V_{\text{dirac}}$, $G_{\text{tunnel}}$ starts to increase (Figs. 1(d) and (e)). In a range of $V_g$ where $-k_B T \lesssim \varepsilon_g < 0$, $G_{\text{thermal}}$ is comparable to $G_{\text{tunnel}}$, so that the non-monotonic decrease in the total resistance can be seen for lower values of $V_{\text{dirac}}$ in Fig. 3(b).

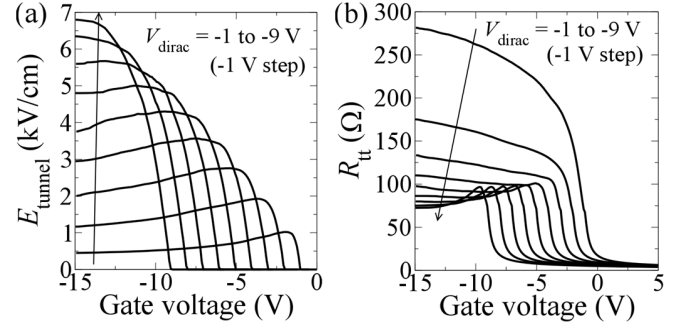

Fig. 3. (a) The electric field at the tunneling point, $E_{\text{tunnel}}$, vs gate voltage with different Dirac voltages and (b) sum of the thermionic-emission and tunneling resistances, $R_{tt}$, vs gate voltage with different Dirac voltages. We set $\alpha_{\text{thermal}} = \alpha_{\text{tunnel}} = 1$. Other parameters used are the same as in Fig. 2.

## C. Extraction of Intrinsic Mobility in Gated Region

Fitting of measured I-V characteristics given as resistances, $R_n$, at gate voltages $V_{gn}$ ($n = 1, 2, \ldots, N$) is done by finding the least squares of the function

$$f = \sum_{n=1}^{N} \left[1 - R(V_{gn})/R_n\right]^2, \quad (11)$$

where $R(V_{gn})$ is calculated as (1), with fitting parameters $V_{\text{dirac}}$, $\alpha_{\text{lin}}$, $U_{\text{imh}}$, $l_{\text{imh}}$, $\alpha_{\text{thermal}}$, $\alpha_{\text{tunnel}}$, and $R_s$. In the fitting demonstrated in the next section, the simulated annealing method was adapted to find the optimal set of the fitting parameters. To reduce the computational cost of the optimization search, we prepared a table of $E_{\text{tunnel}}$ at wide ranges of the gate voltage and Dirac voltage and interpolated it during the search.

Then, the intrinsic mobilities of electrons and holes, $\mu_{\text{int}}^{(e)}$ and $\mu_{\text{int}}^{(h)}$, respectively, in the gated region can be extracted by

$$\mu_{\text{int}}^{(e)} = \sigma_e / en, \quad \mu_{\text{int}}^{(h)} = \sigma_h / ep. \quad (12)$$

Since we assume the symmetry of the electron and hole transports in our model, gate-voltage dependences of the electron and hole mobilities are symmetric with respect to the Dirac voltage. Therefore, we use a term "intrinsic mobility" to refer both of them.

As discussed in subsection III.A, if the nonlinear term in (9) is absent, the total conductivity in (2) is independent of the gate voltage. Similarly, even when the nonlinear term is not too large compared with the linear term, contributions from those terms to the total conductivity and, hence, the resistance can be approximately separable. In other words, the resistance of the gated region is approximately represented as $R_g(V_g) \simeq R_{lin}+R_{inh}(V_g)$, where $R_{lin}$ is the contribution from the linear term and is constant, $R_{inh}$ is that from the nonlinear term. In such a situation, the resistance $R_{lin}$ and the sum of contact resistances and access resistances, $R_s$, cannot be well separated from each other. In turn, the intrinsic mobility cannot be uniquely extracted from an I-V characteristic only, and another complementary measurement is necessary to do so; for example, a direct measurement of $R_s$, or temperature-dependent I-V characteristics. This is not specific to the model developed in this paper but is a fundamental issue for any extraction methods of gate-voltage-dependent intrinsic mobilities of G-FETs, especially those with high-quality graphene channels in which the acoustic-phonon scattering becomes one of major scattering mechanisms.

However, it is possible to extract the range of intrinsic mobility from a single I-V characteristic. First, the lowest value of $\alpha_{lin}$ is equal to $6.58 \times 10^{-3}$ for room-temperature acoustic-phonon scattering. Second, the lowest value of $R_s$ can be roughly estimated from a contact resistivity already reported in literatures (for instance, see [19, 20, 24]) corresponding to the contact materials and the processing techniques that a G-FET under consideration is fabricated with. Those determine the lower and upper bounds of the mobility, respectively. In this case, we first find all the fitting parameters from the fitting, and then we replace either $\alpha_{lin}$ or $R_s$ with its lowest value and find another to give a minimum value of (11).

III. FITTING OF I-V CHARACTERISTIC OF G-FET WITH SiC SUBSTRATE AND SiN GATE DIELECTRIC

Here, we apply the fitting model developed above to a G-FET fabricated and measured in our previous work [11]. It has a graphene channel grown on a C-face 4H-SiC substrate and a SiN gate dielectric deposited by the plasma-enhanced CVD. The geometrical parameters of the G-FET are as follows: the gate length $L_g = 4.4$ μm, thickness $t_g = 45$ nm and relative permittivity $\varepsilon_t = 4.2$ of the gate dielectric, the channel width $W_{ch} = 11$ μm, relative permittivity of the substrate $\varepsilon_b = 9.7$, and the length of access regions $L_a = 100$ nm. The measured relative permittivity of SiN is lower than the known value of SiN in literatures (around 7.5) because deposition conditions were identical with those for growth of large-area, thick SiN passivation layers and not optimized for deposition of thin gate dielectrics. Also, the short access regions were prepared to reduce the access resistance. As the lowest value of the extrinsic resistance, we set it to $R_s = 40$ Ω estimated by $R_s = 2\rho_c W_{ch}$ with the contact resistivity taken from [24].

The assumption (i) of the model mentioned in Sec. II was confirmed by observing a large negative shift of the peak of the I-V characteristic, meaning the channel is highly n-doped, while (ii) was checked by applying the source-drain voltage up to $V_{ds} = 100$ mV. The assumption (iii) is evidently valid, although the short access regions slightly violate (iv). A possible effect of this violation on fitting results shall be discussed later.

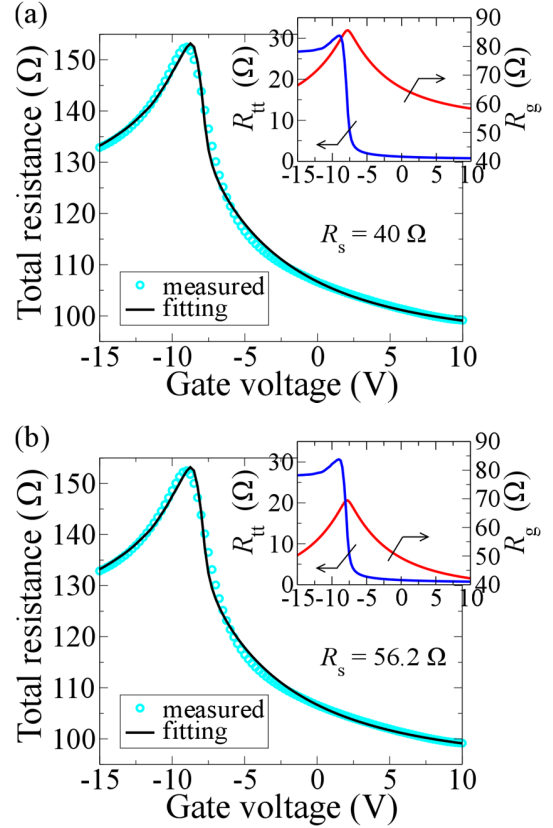

Fig. 4. (a) The total resistance of the G-FET in [11] measured at the drain voltage $V_{ds} = 50$ mV and at room temperature, $T = 300$ K (circles), a fitting curve for it calculated with the lowest value of $R_s = 40$ Ω (solid line), and the decomposition into the gate-voltage dependent portions, $R_g$ and $R_{tt}$ (inset). (b) The same as (a) but for the fitting with the lowest value of $\alpha_{lin} = 6.58 \times 10^{-3}$.

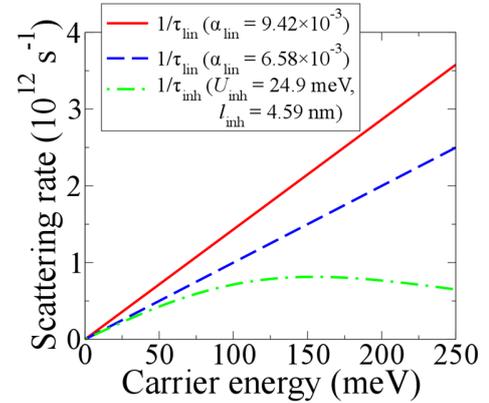

Fig. 5. Carrier-energy dependences of scattering rates $1/\tau_{lin}$ with $\alpha_{lin} = 9.42 \times 10^{-3}$ and $\alpha_{lin} = 6.58 \times 10^{-3}$ (solid and dashed lines, respectively) and $1/\tau_{inh}$ with $U_{imh} = 24.9$ meV and $l_{imh} = 4.59$ nm (dashed-dotted line).

Figure 4(a) shows the total resistance of the G-FET in [11] measured at $V_{ds} = 50$ mV and at room temperature together with a fitting curve for it calculated using the model with $R_s = 40$ Ω. It demonstrates an excellent result of the fitting. Values of the fitting parameters were as follows: $V_{dirac} = -7.82$ V, $\alpha_{lin} = 9.42 \times 10^{-3}$, $U_{imh} = 24.9$ meV, $l_{imh} = 4.59$ nm, $\alpha_{thermal} = 5.33$, $\alpha_{tunnel} = 3.11$. The same fitting curve except setting the lowest value of $\alpha_{lin} = 6.58\times10^{-3}$ is shown in Fig. 4(b). The extracted extrinsic resistance was $R_s = 56.2\ \Omega$. As seen in Figs. 4(a) and (b), the fitting curves do not have visible differences. It illustrates the difficulty of unique extraction of $\alpha_{lin}$ and $R_s$ from a single I-V characteristic and the necessity of finding their ranges (and the intrinsic mobility).

From the extracted Dirac voltage, the doping concentration is obtained as $3.91\times10^{12}$ cm$^{-2}$. It can be seen in Fig. 4(a) that the Dirac voltage does not correspond to the peak of the total resistance, at $V_g = -8.75$ V. This shift takes place because the resistance of the gated region, $R_g$, has a peak at the Dirac voltage but the resistance due to the thermionic emission and interband tunneling, $R_{tt}$, has an abrupt increase there (see the inset in Fig. 4(a)). In general, with an $n$-doped channel, the latter resistance shifts the peak of the total resistance to the "left", and the "left" part of the total resistance shifts up; vice versa with a $p$-doped channel.

Figure 5 shows scattering rates $1/\tau_{lin}$ and $1/\tau_{inh}$ with extracted parameters as functions of carrier energy. Compared to the acoustic-phonon-limited scattering rate, $1/\tau_{lin}$ with $\alpha_{lin} = 6.58\times10^{-3}$, the scattering rates $1/\tau_{lin}$ with $\alpha_{lin} = 9.42\times10^{-3}$ and $1/\tau_{inh}$ are of the same order (the latter is even lower than that). This indicates that in the G-FET under consideration the external scattering sources other than acoustic phonons are remarkably suppressed and the scattering rate is comparable to that by acoustic-phonon scattering.

Values of the extracted fitting parameters for the thermionic emission and interband tunneling, $\alpha_{thermal} = 5.33$ and $\alpha_{tunnel} = 3.11$, which are higher than expected, mean that the analytical expressions (9) and (10) and/or the numerical calculation of $E_{tunnel}$ underestimate those in the real G-FET. For the interband tunneling, neglect of thermal spread of distribution functions when the gated region become $i$-region (i.e., $V_g \sim V_{dirac}$), invalidation of an assumption of smooth $n$-$p$ junctions required for the analytical expression of the tunneling probability [25] to derive (10), and/or inaccuracy of $E_{tunnel}$ due to the neglect of contacts in the numerical calculation can be possible reasons. The last one can be caused by the short access regions in our G-FET. For the thermionic emission, the change in the barrier height by voltage drops at the junctions and the gated region and/or the local heating of electrons at the junctions can cause the increase in the conductance. Although identification of reasons for these discrepancies, as well as correction to the model while keeping computational costs of finding optimal fitting parameters reasonably low, necessitate further investigation, it is out of scope of the paper.

Figures 6(a) and (b) show the gate-voltage-dependent intrinsic mobility ($V_g > V_{dirac}$ for electrons and $V_g < V_{dirac}$ for holes) and the intrinsic transconductance with the voltage drop of 50 mV in the gated region. For the mobility, total concentration ($n$-$p$) corresponding to each gate voltage is also shown. Here, we define the intrinsic transconductance as

$$g_{m,int} = \frac{\partial G_g}{\partial V_g} V_{ds}, \quad (13)$$

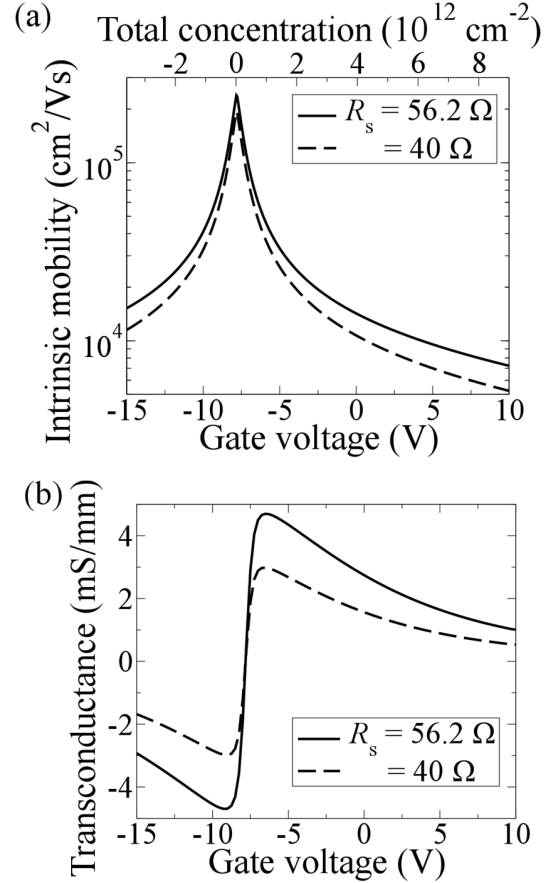

Fig. 6. (a) Lower and upper bounds of extracted intrinsic mobility ($V_g > V_{dirac}$ for electrons and $V_g < V_{dirac}$ for holes), and (b) lower and upper bounds of extracted intrinsic transconductance with the voltage drop of 50 mV in the gated region.

which is the transconductance in the gated region as if all the source-drain voltage is applied to there. As seen in Fig. 6(a), the intrinsic mobility exhibits the maximum of 194,000-239,000 cm$^2$/Vs at the Dirac voltage and of 50,600-63,300 cm$^2$/Vs at the maximum of the intrinsic transconductance. This together with the low extrinsic resistance, 40-56.2 $\Omega$, indicate an excellent performance of the G-FET in [11]. The extracted value of the constant mobility in [11], 101,000 cm$^2$/Vs, is only at the middle between the maximum mobility and the mobility at the maximum intrinsic transconductance, and is insufficient to represent either of the latter values. The gate-voltage-dependent mobility provides more useful information about G-FETs, depending on operating points of the gate voltage in question; e.g., as amplifiers the mobility at maximum transconductance is important. For other applications such as terahertz lasers [26] or plasmonic terahertz devices [27], the concentration-dependent momentum relaxation time is more important to determine the device performances. In either case, the fitting model developed here is a powerful tool to correctly extract intrinsic and extrinsic device parameters of G-FETs, especially the gate-voltage-dependent intrinsic mobility, for evaluation of growth and gate-stack technologies of G-FETs on their performances.

## IV. Conclusion

A fitting model for asymmetric I-V characteristics of G-FETs was developed in order to correctly extract the gate-voltage-dependent intrinsic mobility in the gated region and other intrinsic and extrinsic device parameters. It took into account (i) the energy-dependent momentum relaxation time to describe the gate-voltage-dependent Drude conductivity and, hence, the gate-voltage-dependent mobility, and (ii) the thermionic emission and interband tunneling at $n$-$i$/$i$-$n$ or $n$-$p$/$p$-$n$ junctions formed at the edges of the gated region to describe the asymmetric ambipolarities in I-V characteristics. The following assumptions for device operation and geometry were made: (i) the graphene channel is uniformly doped, (ii) the G-FET operates in the linear regime, (iii) the gated region is sufficiently long, and (iv) the access regions are sufficiently long. Based on these assumptions, the total resistance calculated from the I-V characteristic of the G-FET can be represented as the sum of the resistance of the gated regions, $R_g$, the resistance due to the thermionic emission and interband tunneling at the junctions, $R_{tt}$, and the sum of contact and access resistances, $R_s$. To describe the gate-voltage-dependent $R_g$, it is crucial to include both scattering rates linear and nonlinear in carrier energy, as the absence of the latter causes $R_g$ independent of the gate voltage. Conversely, it was turned out that the former causes a fundamental difficulty in separating the gate-voltage-independent part of $R_g$ from $R_s$. Thus, the intrinsic mobility cannot be uniquely extracted from an I-V characteristic only, and another complementary measurement is necessary. Instead, upper bounds of the mobility was found by assuming only the acoustic phonon scattering for the gate-voltage-independent part of $R_g$, while lower bounds was found by estimating $R_s$ from a lowest value of the contact resistivity already reported in literatures. The model was applied to a top-gated G-FET with a graphene channel grown on a SiC substrate and with SiN gate dielectric we reported previously. We demonstrated that the model can excellently fit the I-V characteristic and can extract its high intrinsic mobility as well as its low extrinsic resistance. It was shown that the Dirac voltage does not correspond to the peak of the total resistance because of an abrupt increase in $R_{tt}$ near the Dirac voltage. These results verified that the model is a powerful tool to correctly extract intrinsic and extrinsic device parameters of G-FETs, especially the gate-voltage-dependent intrinsic mobility, for evaluation of growth and gate-stack technologies of G-FETs on their performances.

## Acknowledgment

This work was financially supported by JSPS Grant-in-Aid for Specially Promoted Research (#23000008). The calculation was partially carried out using the computational resources by the Information Technology Center, the University of Tokyo, and by Research Institute for Information Technology, Kyushu University.

## References

[1] F. Schwierz, "Graphene transistors: status, prospects, and problems," Proc. IEEE, vol. 101, no. 7, pp. 1567-1584 Jul. 2013.
[2] C. Berger et al., "Ultrathin epitaxial graphite: 2D electron gas properties and a route toward graphene-based nanoelectronics," J. Phys. Chem., vol. 108, no. 52, pp. 19912-19916, Dec. 2004.
[3] K. S. Novoselov et al., "Two-dimensional gas of massless Dirac fermions in graphene," Nature, vol. 438, no. 7065, pp. 197-200, Nov. 2005.
[4] S. V. Morozov et al., "Giant intrinsic carrier mobilities in graphene and its bilayer," Phys. Rev. B, vol. 100, no. 1, pp. 016602 1-4, Jan. 2008.
[5] Y. Wu et al., "State-of-the-art graphene high-frequency electronics," vol. 12, no. 6, pp. 3062-3067, May 2012.
[6] Z. Guo et al., "Record maximum oscillation frequency in C-face epitaxial graphene transistors," Nano Lett., vol. 13, no. 3, pp. 942-947, Feb. 2013.
[7] H. Ago et al., "Epitaxial chemical vapor deposition growth of single-layer graphene over cobalt film crystallized on sapphire," ACS NANO, vol. 4, no. 12, pp. 7407-7413, Dec. 2010.
[8] C. Riedl, C. Coletti, T. Iwasaki, A. A. Zakharov, and U. Starke, "Quasi-free-standing epitaxial graphene on SiC obtained by hydrogen intercalation," Phys. Rev. Lett., vol. 103, no. 24, pp. 246804 1-4, Dec. 2009.
[9] S. Tanabe, Y. Sekine, H. Kageshima, and H. Hibino, "Electrical characterization of bilayer graphene formed by hydrogen intercalation of monolayer graphene on SiC(0001)," Jpn. J. Appl. Phys., vol. 51, no. 2S, pp. 02BN02 1-3, Feb. 2012.
[10] L. Banszerus et al., "Ballistic transport exceeding 28 μm in CVD grown graphene," Nano Lett., vol. 16, no. 2, pp. 1387-1391, Feb. 2016.
[11] G. Tamamushi et al., "High carrier mobility graphene-channel FET using SiN gate stack," presented at the Compound Semiconductor Week 2015, Santa Barbara, CA, USA, Jun. 28-Jul. 2, 2015.
[12] S. Kim et al., "Realization of a high mobility dual-gated graphene field-effect transistor with $Al_2O_3$ dielectric," Appl. Phys. Lett., vol. 94, no. 6, pp. 062107 1-3, Feb. 2009.
[13] H. Wang, A. Hsu, J. Kong, D. A. Antoniadis, and T. Palacios, "Compact virtual-source current-voltage model for top- and back-gated graphene field-effect transistors," IEEE Trans. Electron Devices, vol. 58, no. 5, pp. 1523-1533, May 2011.
[14] D. Jiménez and O. Moldovan, "Explicit drain-current model of graphene field-effect transistors targeting analog and radio-frequency applications," IEEE Trans. Electron Devices, vol. 58, no. 11, pp. 4049-4052, Nov. 2011.
[15] S. Frégonèse, M. Magallo, C. Maneux, H. Happy, and T. Zimmer, "Scalable electrical compact modeling for graphene FET transistors," IEEE Trans. Nanotech., vol. 12, no. 4, pp. 539-546, Jul. 2013.
[16] V. Ryzhii, M. Ryzhii, and T. Otsuji, "Tunneling current-voltage characteristics of graphene field-effect transistor," Appl. Phys. Exp., vol. 1, no. 1, pp. 013001 1-3, Dec. 2007.
[17] V. Ryzhii, M. Ryzhii, and T. Otsuji, "Thermionic and tunneling transport mechanisms in graphene field-effect transistors," Phys. Stat. Sol. (a), vol. 205, no. 7, pp. 1527-1533, May 2008.
[18] I. Meric et al., "Curent saturation in zero-bandgap, top-gated graphene field-effect transistors," vol. 3, no. 11, pp. 654-659, Nov. 2008.
[19] S. Russo, M. F. Craciun, M. Yamamoto, A. F. Morpurgo, and S. Tarucha, "Contact resistance in graphene-based devices," Physica E, vol. 42, no. 4, pp. 677-679, Feb. 2010.
[20] K. Nagashio, T. Nishimura, K. Kita, and A. Toriumi, "Contact resistivity and current flow path at metal/graphene contact," Appl. Phys. Lett., vol. 97, no. 14, pp. 143514 1-4, Oct. 2010.
[21] K. Takase, S. Tanabe, S. Sasaki, H. Hibino, and K. Muraki, "Impact of graphene quantum capacitance on transport spectroscopy," Phys. Rev. B, vol. 86, no. 16, pp. 165435 1-8, Oct. 2012.
[22] E. H. Hwang and S. Das Sarma, "Acoustic phonon scattering limited carrier mobility in two-dimensional extrinsic graphene," Phys. Rev. B, vol. 77, no. 11, pp. 115449, Mar. 2008.
[23] F. T. Vasko and V. Ryzhii, "Voltage and temperature dependencies of conductivity in gated graphene," Phys. Rev. B, vol. 76, no. 23, pp. 233404 1-4, Dec. 2007.
[24] J. A. Robinson et al., "Contacting graphene," Appl. Phys. Lett., vol. 98, no. 5, pp. 053103 1-3, Jan. 2011.
[25] V. V. Cheianov and V. I. Fal'ko, "Selective transmission of Dirac electrons and ballistic magnetoresistance of $n$-$p$ junctions in graphene," Phys. Rev. B, vol. 74, no. 4, pp. 041403(R) 1-4, Jul. 2006.
[26] T. Otsuji, S. Boubanga Tombet, A. Satou, M. Ryzhii, and V. Ryzhii, "Terahertz-wave generation using graphene: toward new types of terahertz lasers," IEEE J. Sel. Topics Quantum Electron., vol. 19, no. 1, pp. 8400209

1-9, Jan.-Feb. 2013.
[27] T. Otsuji and M. Shur, "Terahertz plasmonics," IEEE Microw. Mag., vol. 15, no. 7, pp. 43-50, Nov.-Dec. 2014.

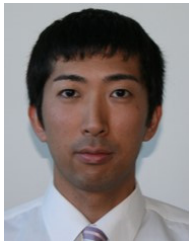

**Akira Satou** received his B.S., M.S., Dr. of Computer Science degrees from University of Aizu, Japan, in 2003, 2005, and 2008, respectively. He was an assistant lecturer at University of Aizu in 2008 and 2009. Since 2010, he has been an assistant professor at Research Institute of Electrical Communication, Tohoku University, Japan. His research interests are theoretical study of carrier transport in heterostructure two-dimensional electron gas and graphene for terahertz applications. He is the author and co-author of more than 50 papers in refereed journals.

Dr. Satou is a Senior Member of IEEE and a member of American Physical Society (APS), the Institute of Electronics, Information, and Communication Engineers of Japan (IEICE), and the Japan Society of Applied Physics (JSAP).

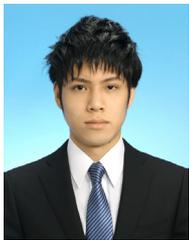

**Gen Tamamushi** received the B.S. degree in information and electronic system engineering in Sendai National College of Technology, Sendai, Japan, in 2014. He is currently pursuing the M.S. degree in electrical and communication engineering at Tohoku University.

Mr. Tamamushi is a student member of the Japan Society of Applied Physics (JSAP).

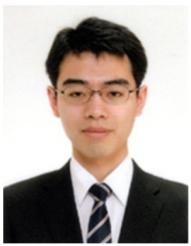

**Kenta Sugawara** received the B.S. degree in electrical and communication engineering in Tohoku University, Sendai, Japan, in 2014. He is currently pursuing the M.S. degree in electrical and communication engineering at Tohoku University. His current research interest includes semiconductor fabrication in small size devices and applications using graphene channel FETs.

Mr. Sugawara is a student member of the Institute of Electronics, Information, and Communication Engineering (IEICE) of Japan, and the Japan Society of Applied Physics (JSAP).

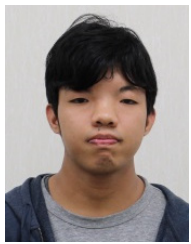

**Junki Mitsushio** received the B.S. degree in physical engineering from Tokyo University of Agriculture and Technology, Tokyo, Japan, in 2015. He is currently pursuing the M.S. degree in electrical and communication engineering at Tohoku University.

Mr. Mitsushio is a student member of the Japan Society of Applied Physics (JSAP).

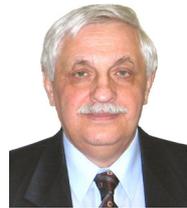

**Victor Ryzhii** (F'04, SM'93) obtained his Ph.D and Doctor of Science (Habilitation) degrees from Moscow Institute of Physics and Technology, Russia, in 1970 and 1975, respectively.

From 1970 to 1993 he worked in academic and industrial institutions in Russia. In 1993 he joined the University of Aizu, Japan, where he worked as a professor until 2012. Currently, he is Professor Emeritus of the University of Aizu and a Visiting Professor of the Research Institute of Electrical Communication, Tohoku University, Japan. His research activity is associated with physics and computer modeling of low-dimensional semiconductor heterostructures and electronic, optoelectronic, and terahertz devices based on nanostructures, including graphene-based devices. He is authored and co-authored more than 300 peer-reviewed journal publications, numerous conference papers, and 11 patents.

Dr. Ryzhii is a Fellow of IEEE, the Institute of Electrical and Electronics Engineers (IEICE), Japan, and the American Physical Society (APS), a Corresponding Member of the Russian Academy of Sciences, and a member of the Japan Society of Applied Physics (JSAP)

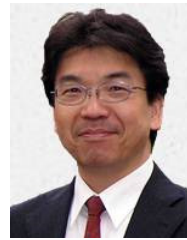

**Taiichi Otsuji** (F'14, SM'13, M'91) was born in Fukuoka, Japan, on September 5, 1959. He received the B.S. and M.S. degrees in electronic engineering from Kyushu Institute of Technology, Fukuoka, Japan, in 1982 and 1984, respectively, and the Ph.D. degree in electronic engineering from Tokyo Institute of Technology, Tokyo, Japan, in 1994.

From 1984 to 1999 he worked for NTT Laboratories, Kanagawa, Japan. In 1999 he joined Kyushu Institute of Technology as an associate professor, being a professor in 2002. Since 2005 he has been a professor at the Research Institute of Electrical Communication, Tohoku University, Sendai, Japan. His current research interests include terahertz electronic and photonic materials/devices and their applications. He is authored and co-authored more than 210 peer-reviewed journals.

Dr. Otsuji is a Senior Member of the Optical Society of America (OSA), and a member of the Materials Research Society (MRS), the International Society for Optics and Photonics (SPIE), the Institute of Electronics, Information, and Communication Engineers of Japan (IEICE) and the Japan Society of Applied Physics (JSAP). He was the recipient of the Outstanding Paper Award of the 1997 IEEE GaAs IC Symposium. He has been served as a Distinguished Lecturer, Electron Device Society, IEEE, since 2012.